\documentclass[prb,twocolumn,showpacs,preprintnumbers,amsmath,amssymb]{revtex4}

\usepackage{graphicx}
\usepackage{amsmath}

\begin{document}

\title{
     	Reversal Modes of Simulated Iron Nanopillars in an Obliquely Oriented Field
      }

\author{
	S.H. Thompson 
	}

\affiliation{
	Department of Physics,	School of Computational Science,\\
	and Center for Materials Science and Technology,\\
	Florida State University, Tallahassee, FL 32306-4350, USA
	}
\author{
	G. Brown
	}
\affiliation{
	Center for Computational Sciences,~Oak Ridge National Laboratory,\\
	Oak Ridge, TN 37831-6164, USA 
	}
\author{
	P.A. Rikvold
	}
\affiliation{
	School of Computational Science, Center for Materials Science and Technology, and\\
	Department of Physics, Florida State University, Tallahassee, FL 32306-4120, USA\\	
	and National High Magnetic Field Laboratory, Tallahassee, FL 32310-3706, USA\\
	}	

\date{\today}

\begin{abstract}
Stochastic micromagnetic simulations are employed to study switching in 
three-dimensional magnetic nanopillars exposed to highly misaligned fields.
The switching appears to proceed through two different decay modes,
characterized by very different average lifetimes and different average
values of the transverse magnetization components.
\end{abstract}

\pacs{
	75.75+a, 61.46+w, 64.60.Qb
	}
	
\maketitle

\section{Introduction}
Nanoparticles with strong uniaxial anisotropy are frequently employed in technologies where the orientation of the particle
magnetization corresponds to a binary digit.
The switching process of the magnetization in an applied field therefore becomes a technologically important process, 
especially for fields that vary on a timescale of nanoseconds.
Micromagnetic simulations have been essential for understanding properties of systems that evolve
over such short time scales.

\begin{figure}[tb]
\centering\includegraphics[scale=0.40]{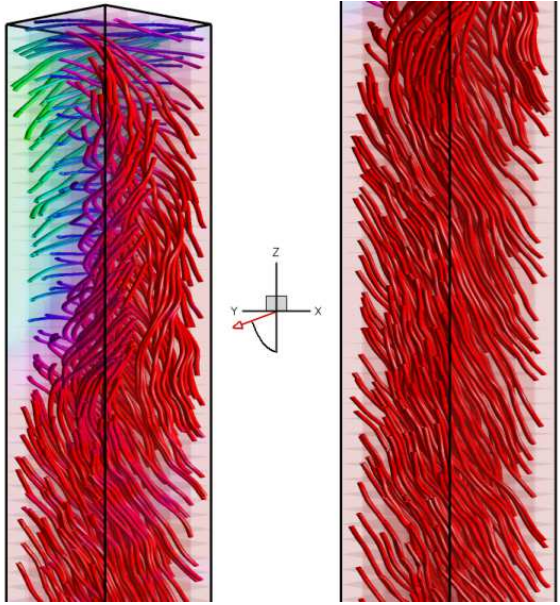}
\caption[] {
	Streamline rendering of a nanopillar before reversing to the stable state.
	$M_z$ is denoted by the color of the streamlines: $+z$ is dark gray, $-z$ is light gray (red/green for color online).
	The applied field represented by the arrow in this right-handed coordinate system is oriented 75$^\circ$ from the $-z$-axis,
	parallel to the $zx$ plane (pointing to the left and out of the page).  The left part shows the top of a pillar;
	the right part, the same pillar further down the easy axis.
	}
\label{fig:Streamline}
\end{figure}

In this paper we examine a model nanopillar where the anisotropy is due solely to its shape,
and subject it to a reversal field that is highly misaligned with respect to the easy axis of
the pillar.
Using parameters consistent with bulk iron, we find that several interesting features emerge as the magnetization reverses
from an unstable configuration, saturated opposite the applied field, to the true, stable state.
Most noticeable is that this highly misaligned system possesses two distinct decay paths that evolve from
the same initial conditions.
Figure~\ref{fig:Streamline} is an example snapshot of a slowly reversing decay path, at a time when the average $z$ component $M_z$ of the system magnetization 
corresponds to a relatively flat region of the free-energy landscape.
As seen in other simulations with applied fields along the easy axis of the pillar,\cite{BROWN_PRB01} the end caps display a region of high vorticity
due to pole avoidance at the end faces of the pillar.
This is shown in the left part of Fig.~\ref{fig:Streamline}.
With an applied field nearly parallel to the easy axis, the flux lines continue down the volume of the pillar, mostly parallel to the easy axis.
The highly misaligned case discussed in this paper, however, exhibits magnetic flux lines that penetrate faces parallel to the $yz$ plane 
near the mid-section, as shown in the right part of Fig.~\ref{fig:Streamline}.

Many nanomagnets have lateral dimensions on the order of the exchange length
and can be sufficiently represented by the Stoner-Wolhfarth mechanism of coherent rotation.\cite{HERTEL_KIRSCHNER04}
Larger systems that span several exchange lengths such as the one discussed in this paper, however, are not adequately described by coherent rotation
and consequently require the application of full micromagnetics simulations to allow the proper dynamics to emerge.
To investigate the dynamics of these nanopillars under the influence of an applied field, we use a 
micromagnetic simulation performing numerical solutions of the Landau-Lifshitz-Gilbert (LLG) equation.\cite{WFBROWN, AHARONI}
For our model, the pillar's real dimension of 10~nm~$\times$~10~nm~$\times$~150~nm are mapped onto a 6~$\times$~6~$\times$~90
computational lattice, and the LLG equation is evaluated at each site in the presence of a local field, $\vec{H}(\vec{r_i})$.  
The net saturation magnetization $M_s$ is introduced in the LLG equation,
\begin{equation}
\frac{\rm{d}\it{\vec{M}(\vec{r_i})}}{\rm{d\it{t}}} = \frac{\gamma_0}{1+\alpha}\left(\vec{M}(\vec{r_i})\times 
\left[\vec{H}(\vec{r_i})-\frac{\alpha}{M_s}\vec{M}(\vec{r_i})\times \vec{H}(\vec{r_i})\right]\right),
\label{eq:M}
\end{equation}
as a constant magnitude of every unit magnetization vector $\vec{M}(\vec{r_i})$ in the lattice.
Also included in the equation are the constant electron gyromagnetic ratio $\gamma_0 = 1.67\times10^{7}$ Hz/Oe and a
phenomenological damping parameter, whose value we take to be $\alpha = 0.1$.
The local field $\vec{H}(\vec{r_i})$ that determines the dynamics at each site $i$ is a sum of all participating fields,
including applied, dipole, exchange, and a stochastic thermal field with a Gaussian distribution that obeys the 
fluctuation-dissipation theorem.
In this simulation the applied field was set to 3160 Oe at an angle of 75$^\circ$ to the easy axis, which is near the coercive field for the
slow decay mode.
The temperature was set to 20 K.

Initially, the nanopillar was allowed to relax in the presence of an unreversed field before sinusoidally reversing
the field $180^\circ$ to drive the reversal process.
Measurements were taken at regular intervals during the simulation as the LLG equation was temporally integrated using an Euler integration scheme.

\section{Results and Conclusions}

\begin{figure}[tb]
\centering\includegraphics[scale=0.30]{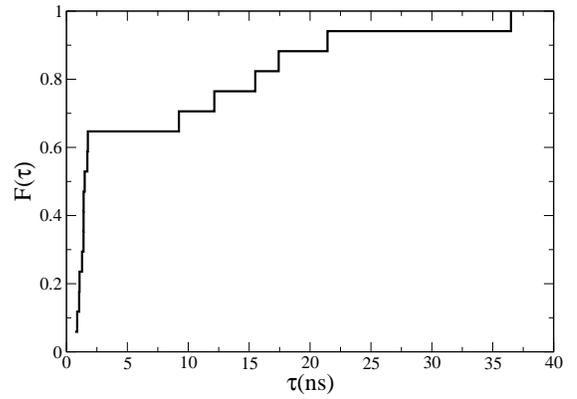}
\caption[] {
	Cumulative distribution of the lifetimes in nanoseconds.  The separation in a fast and a slow mode is clearly seen.
	}
\label{fig:CD}
\end{figure}

\begin{figure}[tb]
\centering\includegraphics[scale=0.30]{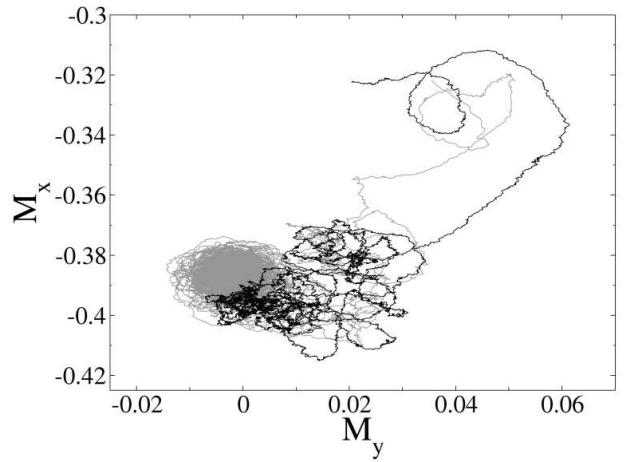}
\caption[] {
	Plot of the $M_x$$M_y$ phase plane for the slow (gray) and fast (black) mode.  
	Simulations begin close to center of the graph, evolve through the metastable
	region, and end near the top right corner.
	}
\label{fig:MxMy}
\end{figure}

The first evidence for the existence of two separate decay modes in the nanopillar comes from the distribution of the lifetimes $\tau$, determined by the time
it took the $z$-component of the total magnetization, $M_z$, to reach a value of~$-0.75$.
Lifetimes of decay processes such as this one are strongly determined by the shape of the free-energy landscape that the system 
evolves through.
For free-energy barriers that are much larger than the temperature ($\gg k_BT$), an exponential distribution of the lifetimes
should be observed.
Regions with negligible barriers lead to a Gaussian distribution of the lifetimes, corresponding to perturbations around a deterministic behavior.
In Fig.~\ref{fig:CD} the cumulative distribution for all our 17 simulations is shown.
The graph is clearly divided into two regions.
The first region corresponds to a rapid increase in the distribution for times less than about $2$ nanoseconds, and the second to a much slower increase that is spread across
tens of nanoseconds.
For the fast switches, evidence presented below suggests there is no free energy barrier. 
In that case, we expect the fast switching times to have an approximately Gaussian distribution.
It is not clear what form the distribution for the slower switches has.

Another useful method that reveals the different reversal modes consists of recording the total magnetization 
(which we normalize to be unity when all spins are aligned)
of the nanopillar at regular time steps and constructing a phase plot of the evolution.
Two examples, with $\tau=1.6$ and 17.6 nanoseconds, respectively, are shown in Fig.~\ref{fig:MxMy} for the $x$ and $y$ components of the 
total magnetization.
Both configurations begin their evolution near the middle of the plot and evolve through their respective transition states, ending up close
to the top right corner.
There is a difference in the regions where these two examples spend most of the time before reversal, and this difference is seen for all simulations.
This behavior is consistent with a local minimum in the free energy for the slow mode, and a much shallower (or nonexistent) minimum for the fast mode.
However, from the projective dynamics evidence to be presented next, it is not clear that a true minimum exists, even for the slow mode.

\begin{figure}[tb]
\centering\includegraphics[scale=0.30]{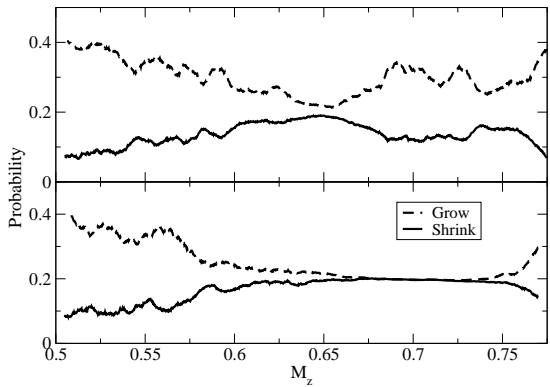}
\caption[] {
	Growth and shrinkage probabilities distinguishing the two decay modes in the simulation.
	The location and depth of the metastable well in the free energy depend
	on the reversal mode.  Upper panel: fast mode.  Lower panel: slow mode.
	}
\label{fig:PD}
\end{figure}

Projective dynamics\cite{KOLESIK, BROWN_PB03, THOMPSON} gives information contrasting the two modes in this nanopillar simulation.
By selecting a slow variable, growth and shrinkage probabilities are accumulated in bins
along that variable and can be used to further elucidate properties of the free-energy landscape.
Specifically, the difference between the growth and shrinkage probabilities is a measure of the shape of the free energy along that variable.
In this instance, $M_z$ was chosen for binning, and transition probabilities were constructed based on the number of
times $M_z$ jumped to an adjacent bin.
The results of the projective dynamics analysis are shown in Fig.~\ref{fig:PD}.
For the slower reversal mode, the growth and shrinkage probabilities practically coincide in the region around $M_z \thickapprox 0.7$.
This indicates that the free energy is almost flat in this region and that there is no strongly preferred direction of motion for $M_z$.
In comparison, the growth probabilities always exceed the shrinkage probabilities for the fast mode, indicating a
tendency for the magnetization to continuously evolve toward the stable state.
Since neither of these graphs exhibits an area where the shrinkage probability exceeds the growth probability, it is 
possible that a significant barrier in the free energy does not exist.
It is also possible that $M_z$ is not a good choice for the slow variable, e.g. it may not coincide well with the reversal path.
The statistics that were collected to construct the projective dynamics plots were sorted based on the lifetime
of the particular run.
Due to the approximate classification of the modes, it is likely that there is some mixing between the data in these plots, although not
enough to change the general features observed here.

In conclusion, finite-temperature simulations were performed on iron nanopillars in a strongly misaligned applied 
field.
Sophisticated visualization tools were used to give a detailed picture of the magnetization configurations during the decay.
Two distinct reversal modes were found, despite identical initial conditions.  
Three different methods were used to characterize the differences between these two modes: the distribution of the lifetimes, phase plots
of the magnetization transverse to the easy axis, and projective dynamics.
Still, our results do not yet provide a very clear picture of the shape of the free-energy landscape.
Fully characterizing the free energy will be important for designing nanomagnets that switch in a reliable manner under
given experimental conditions, and will be a topic of future research.

\begin{acknowledgments}

This work was supported in part by U.S. NSF grant No.~DMR-0120310 and by Florida State University through the Center
for Materials Research and Technology and the School of Computational Science.

\end{acknowledgments}

\end{document}